\documentclass[letterpaper, 10 pt, conference]{ieeeconf}  

\IEEEoverridecommandlockouts                              

\usepackage[pdftex]{graphicx}
\usepackage{amsmath} 

\title{\LARGE \bf
Reslicing Ultrasound Images for Data Augmentation\\ and Vessel Reconstruction
}

\author{Cecilia G. Morales, Jason Yao, Tejas Rane, Robert Edman, Howie Choset, Artur Dubrawski \\
\thanks{*This work was supported by U.S.\ Department of Defense contracts W81XWH-19-C0083, W81XWH-19-C0101, and W81XWH-19-C-0020}
\thanks{Authors are with the Robotics Institute, Carnegie Mellon University, Pittsburgh, PA, USA
{\tt\small \{cgmorale, jlyao, tejasr, redman, choset, awd\}@andrew.cmu.edu}}
        }
        
\usepackage{todonotes}
\setuptodonotes{fancyline, color=blue!30}
\usepackage{svg}
\usepackage{booktabs}
\usepackage{tabularx}

\begin{document}

\maketitle
\thispagestyle{empty}
\pagestyle{empty}

\begin{abstract}
Robot-guided catheter insertion has the potential to deliver urgent medical care in situations where medical personnel are unavailable. However, this technique requires accurate and reliable segmentation of anatomical landmarks in the body. For the ultrasound imaging modality, obtaining large amounts of training data for a segmentation model is time-consuming and expensive. This paper introduces RESUS (\underline{RES}licing of \underline{U}ltra\underline{S}ound Images), a weak supervision data augmentation technique for ultrasound images based on slicing reconstructed 3D volumes from tracked 2D images.  This technique allows us to generate views which cannot be easily obtained in vivo due to physical constraints of ultrasound imaging, and use these augmented ultrasound images to train a semantic segmentation model. We demonstrate that RESUS achieves statistically significant improvement over training with non-augmented images and highlight qualitative improvements through vessel reconstruction.

\end{abstract}
\section{INTRODUCTION}
\label{sec:introduction}
Adequate vascular access in trauma patients prior to decompensation of physiological processes is essential for a patient’s survival. Vascular access is of paramount importance to provide a route for anesthesia or other medications, resuscitative fluids, intravenous contrast for diagnostic procedures, and it may be a useful adjunct for measuring physiological parameters during primary resuscitation \cite{Verhoeff_Saybel_Mathura_Tsang_Fawcett_Widder_2018, buckenmaier_mahoney_2017}. An example where the provision of fluids is contingent on obtaining suitable vascular access to the patient’s venous system is hemorrhaging, the leading cause of death in severely injured trauma patients: it accounts for 31\% of deaths in the first hour after injury \cite{Verhoeff_Saybel_Mathura_Tsang_Fawcett_Widder_2018}. Clinical management involves achieving hemostasis by replacing lost intravascular volume with fluids and blood, and treating coagulopathy \cite{statPearls}. Other treatments use the insertion of a vascular catheter for Resuscitative Endovascular Balloon Occlusion of the Aorta (REBOA) via the femoral artery to prevent significant loss of blood or extracorporeal membrane oxygenation (ECMO).

For large volume resuscitation, the insertion of a central venous line is potentially lifesaving since it enables rapid administration of high volumes of isotonic fluids and medications that would be caustic to peripheral veins \cite{statPearls}. Taking into account the rapid deterioration in the health of a patient, it is vital that a patient is treated within a short period of time in order to save the patient’s life. However, in situations where the patient is in a remote location such as a soldier on the battlefield, or in massive casualty scenarios, access to appropriate treatment resources can be limited.

While catheterization is a standard procedure in clinical practice, robot-guided catheter insertion could enable necessary medical care to the patient within a short period of time where medical personnel are not yet readily available. Given the important role that femoral blood vessels have in the resuscitation of a patient, it is important to identify them inside the leg. To be able to determine the proper robot needle insertion location autonomously, a suitable medical imaging modality is necessary. 

Medical imaging involves non-invasive procedures aimed to visualize anatomical features within the body. Some of the most common imaging techniques in clinical practice are X-ray, computer tomography (CT), ultrasound (US), magnetic resonance (MR), and positron emission tomography (PET). This work utilizes ultrasound as it allows real-time imaging, easy portability, no need for intravenous contrast agents, no ionizing radiation, and low-costs.

Ultrasound imaging, also called sonography, uses sound waves to capture internal images of the body. An instrument called a transducer emits high-frequency sound, inaudible to human ears, and records the echoes as the sound waves bounce back to determine the size, shape, and consistency of soft tissues and organs \cite{CypressDiagnostic}. 

\begin{figure}[tp]
  \centering
  \includegraphics[width=0.75\columnwidth]{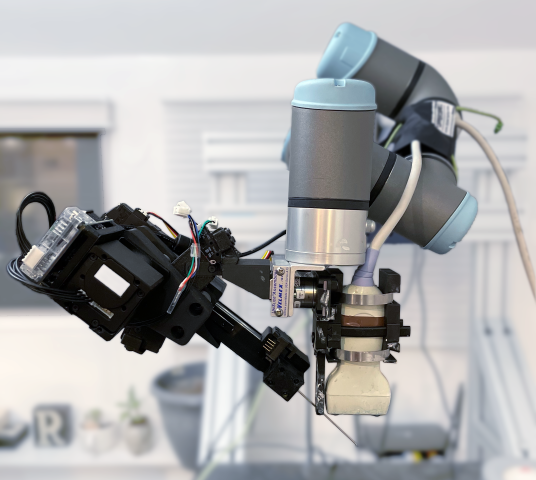}
  \caption{Our Mechanical Ultrasound Scanning System, which consists of a linear array ultrasound transducer attached to the end-effector of a 6-DoF Universal Robots UR3e serial manipulator.}
   \vspace{-15pt}
  \label{figurelabel}
\end{figure}

By attaching an ultrasound transducer to the robot's end-effector and utilizing the method described in \cite{10.1109/ICRA46639.2022.9812410}, we can scan the femoral region to identify the exact location of the vessels. Using a robotic ultrasound system ensures accuracy, stability, maneuverability, and tracking the location of the transducer during ultrasound image acquisition.

Nonetheless, automatically segmenting anatomical landmarks in the body such as arteries or veins is a challenging task. One must consider the noise and artifacts in images, inter-patient and machine variability, as well as natural and injury-inflicted anatomical differences. Acquiring comprehensive sets of training data for all these scenarios would be time-consuming, difficult and expensive, since only professionally trained personnel are able to adjudicate them.


In this work, we introduce RESUS (\textbf{RES}licing of \textbf{U}ltra\textbf{S}ound Images), a weak supervision data augmentation technique for ultrasound images. Specifically, the method is designed to reconstruct 3D volumes made up of tracked 2D ultrasound images. The volume can then be resliced to reveal 2D images in planes that would not be accessible due to the physical restrictions of the scanning process, i.e., a plane parallel to the surface of the skin. These new reconstructed ultrasound images are then used to train a semantic segmentation U-Net network \cite{10.1007/978-3-319-24574-4_28} such that it is able to generalize to different anatomical variations and slight altercations, artifacts, and varying image resolutions.

We collect ultrasound images with a robotic arm through a series of experiments on a medical imaging phantom as well as real-world data. We compare our method with the predictions of a U-Net trained on non-augmented images, a U-Net pretrained on ImageNet, and a U-Net trained on random augmentation images. RESUS achieves statistically significant, respectively 8\%, 5\%, and 23\% increases in Intersection over Union (IoU) score, also known as Jaccard Index, in segmenting images.

The main contributions of our work are: 1) Developing a novel method that augments ultrasound images to train on new imaging perspectives which are otherwise inaccessible due to the physical constraints of the system; 2) Creating a  novel volume reconstruction technique to adapt to the data collected; 3) Evaluating the effectiveness of our method both qualitatively and quantitatively in increasing the IoU score; 4) Achieving better vessel reconstructions affordably using weak supervision on real-world data; 5) Increasing the availability of data and easing the labelling task.

\section{BACKGROUND AND RELATED WORK}
\label{sec:relatedWorks}
\begin{figure*}
  \centering
  \def\svgwidth{\textwidth}
  \includegraphics[width=0.99\textwidth]{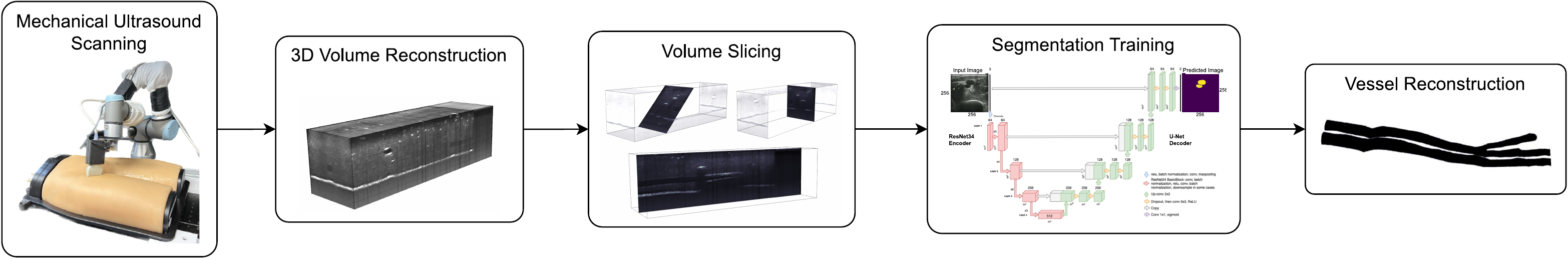}
  \vspace{-5pt}
  \caption{Flow diagram of the proposed method. We collect 2D ultrasound images from a linear array ultrasound transducer attached to the end-effector of the 6-DoF Universal Robots UR3e serial manipulator. A 3D volume is reconstructed from these images, and then resliced to generate frames, some of which which would have been inaccessible during scanning. These frames are then added to the training data to obtain the segmentation model, which identifies the vessels. The segmented vessels inform vessel reconstruction needed to find the optimal needle insertion location.}
  \label{figurelabel}
  \vspace{-15pt}
\end{figure*}

\subsection{Ultrasound Systems}
The three main types of ultrasound (US) are the 3D ultrasound system, and the mechanical and freehand scanning system using a 2D ultrasound probe. In the 3D ultrasound system, data is directly acquired by a series of dedicated 3D probes with an oscillating mechanism that sweeps a predefined region of interest \cite{Huang2017-ii}. Although, it allows a whole volume of data to be acquired at a time, it is expensive and not commonly available. Furthermore, the size of the acquired volume is limited by the dimensions of the transducer which is incapable of scanning large volume organs. The freehand scanning method with a position tracker is comparatively cheaper and more flexible, but the frames acquired typically are irregularly spaced and highly sparse.

Our approach uses the mechanical scanning system with a conventional linear array transducer. 
During ultrasound image acquisition, the transducer is maneuvered under the computer control to collect regularly spaced ultrasound images, and the pose of the scanning head is recorded synchronously with the image \cite{Huang2017-ii}. Although, such systems are relatively costly, they allow for more reliable scanning and tracking, and the limited scanning range is enough to perform a simple linear scan along the leg of a patient.

\subsection{Volume Reconstruction Strategies}
Volume reconstruction starts by obtaining the ultrasound images from a linear array probe and their corresponding collection poses. The two main reconstruction methods are pixel based and voxel based.

Pixel nearest neighbor (PNN) is the most common pixel based reconstruction method. It proceeds traveling across each pixel of the acquired 2D ultrasound images, and the nearest voxel in the 3D reconstructed volume is filled with that pixel value \cite{recon-freeh}. Although PNN results in loss of important information from the 2D images, it can still be favorable for 3D reconstruction because of its simplicity and relatively low computation costs~\cite{3dUS-survey}. 

Voxel based methods are used to reconstruct the 3D volume by traveling across each voxel in the volume grid and gathering the pixel values from input 2D ultrasound frames, and computing the voxel values~\cite{3dUS-survey}. The most common ways of computing the voxel values include voxel-nearest neighbor (VNN) and distance-weighted (DW). The VNN selects the nearest pixel value from a set of 2D frames to be put on the voxel, whereas the DW takes a weighted mean of local neighborhood pixels by the inverse distances between the pixel and the voxel \cite{3dUS-survey}. The VNN is able to preserve the original texture patterns from the 2D frames, but it tends to generate large reconstruction artifacts and preserve the speckle noise from corrupted frames. The DW is able to suppress the speckle noise, but it smooths out the 3D reconstructed volume, causing loss of important information.

 
Since our training data is dense and reconstruction speed is not a high priority, we use the squared-distance-weighted (SDW) reconstruction algorithm from Huang et al.~\cite{HUANG2005153}, an extension of the DW method that aims to reduce the smoothing effect in the reconstructed volume. 

\subsection{Semantic Segmentation using Deep Neural Networks}
U-Net~\cite{unet} is one of the deep learning techniques most widely used  for analyzing medical images to identify organs and lesions. Manual segmentation of images implies a heavy workload for doctors, and it can introduce bias if it involves their subjective opinions.
U-Net is designed primarily for image segmentation \cite{unet-variants}. The basic structure of the U-Net architecture consists of two paths: the encoder, which provides classification information, and the decoder, which helps in localizing the classification information. This structure meets the requirements of medical image segmentation with its fast training speeds, and ability to yield reliable models using small amounts of training data. Various strategies are applied to modify the U-Net structure, to address different segmentation problems, and it has improved in the areas of application range, feature enhancement, training speed optimization, training accuracy, feature fusion, small sample training set, and generalization improvement \cite{unet}.

Another approach adopted by \cite{mit} used the You Only Look Once (YOLO) v3-Tiny network to detect and compute bounding boxes on arteries, veins, and bifurcations. The network was first trained on the ImageNet \cite{imagenet} database, which consists of millions of natural images. Then, to adapt the network to ultrasound images, transfer learning was performed by retraining the network using a database of annotated ultrasound images of the porcine femoral region. Furthermore, affine transformations and random intensities were used as data augmentation techniques to improve the generalization performance.

\subsection{Augmentations}
When working with deep neural networks for semantic segmentation, data augmentation is a relatively simple and commonly used method for generalizing across unseen domains and to account for invariances. Many data augmentation methods for medical images involve simple image transformations, such as flips, skips, skews and blurs. However these augmentations do not seem to have much of an impact in the context of ultrasound images. 

A novel data augmentation technique for ultrasound images was developed by \cite{ed-aug}, where a 3D U-Net was trained adaptively to generalize over different anatomical variations. The augmentation module, an agent feed forward neural network, is responsible for generating the synthetic images for further training \cite{ed-aug}. This method was able to enhance the generalizability of the 3D U-Net, but at the cost of high memory usage and long computation time, therefore not being feasible for real-time field applications, or large amounts of data. Another limitation of this method was that it could miss image deformations in areas of the image not covered by the segmentation ground-truth labels.

Our method improves the aforementioned approaches by reslicing a 3D reconstructed volume to create augmented data which is more intuitive and interpretable than performing random transformations for data augmentation or applying transfer learning when the domain images are highly distinct. This allows us to re-generate unseen views in the form of raw images, such as a longitudinal slice where all the raw images were transverse images, and create slight variations between vessels that could be seen in a new patient. By generating a variety of reconstructed views we increase the volume and comprehensiveness of training data for the segmentation network, as the single reconstructed volume can generate many different altercations of the vessels and resulting images.

\section{METHOD}
\label{sec:method}
In our approach, a robot with an ultrasound probe in its end-effector scans the femoral region of the leg and reconstructs a volume that is then sliced to reveal 2D images that would have not been accessible for the ultrasound due to the physical restrictions of the scanning process. This produces a diverse set of views of the vessel that help generalize to different patients. Those images are used to train a vessel segmentation model. Once the vessels are segmented, they can be reconstructed and used to find appropriate needle insertion sites. 

\subsection{Dataset}
Our method was first tested on a medical imaging phantom, CAE Blue Phantom anthropomorphic gel model. Images were scanned from both the right side and left side of the phantom. Once the pipeline was ready, we transitioned to experiments in a surgical environment at the University of Pittsburgh Medical Center (UPMC) with live pigs under anesthesia. The experiments involving live animals have been conducted in accordance with the Institutional Animal Care and Use Committee (IACUC) protocol approved by the cognizant authority. We collected ultrasound images of the femoral arteries and veins from eight different pigs, 500 to 2070 images each, using the 6-DoF Universal Robot UR3e serial manipulator to move the probe used for scanning. The probe used was the Fukuda Denshi portable point-of-care scanner (POCUS) with 5-12MHz linear transducer with a maximum depth of 5cm and 10cm, respectively. A total of 14,207 images were collected and labelled by expert clinicians using the Computer Vision Annotating Tool (CVAT) \cite{boris_sekachev_2020_4009388}. Robot Operating System (ROS) \cite{ros} was used to record the poses of the robot and concurrent ultrasound images.

\subsection{Volume Definition}
We first define the voxel volume where the reconstruction will take place. From the scanned images, we pick a start and end frame to exclude less useful data such as stationary images and extraneous movement. From the resulting subset of image frames, all the pixels in the images are treated as points in a point cloud and transformed to their position in space according to the transforms from ROS.

We then assume that the resulting point cloud is shaped like a parallelepiped and choose one of its corners to be the new origin of the voxel volume and vectors to its three adjacent vertices become the new basis vectors. We perform a change of coordinates on all the points in the point cloud to the new volume coordinates, resulting in all of the pixels that would contribute to the reconstruction ending up in the unit cube. We then scale each dimension up by the number of voxels desired in the final volume. We assume the point cloud is shaped like a parallelepiped since scanning vessels in the leg need a relatively linear motion without many complex movements.

The main challenge for this volume definition method is finding the relevant vertices of the parallelepiped. We first fit planes to each side of the parallelepiped using a modified RANSAC \cite{ransac} algorithm on the subset of the point cloud corresponding to each face, which helps reduce the effects of extraneous movements on the overall volume definition. The algorithm repeatedly samples 3 points randomly from the point cloud and defines a plane through them. It then computes the distance of that plane to all the points in the point cloud and counts the number of inlier points within a certain distance threshold. The plane that has the greatest number of inliers after a set number of iterations is chosen as the best-fit plane.

\begin{figure}[tp]
  \centering
  \includegraphics[width=0.99\columnwidth]{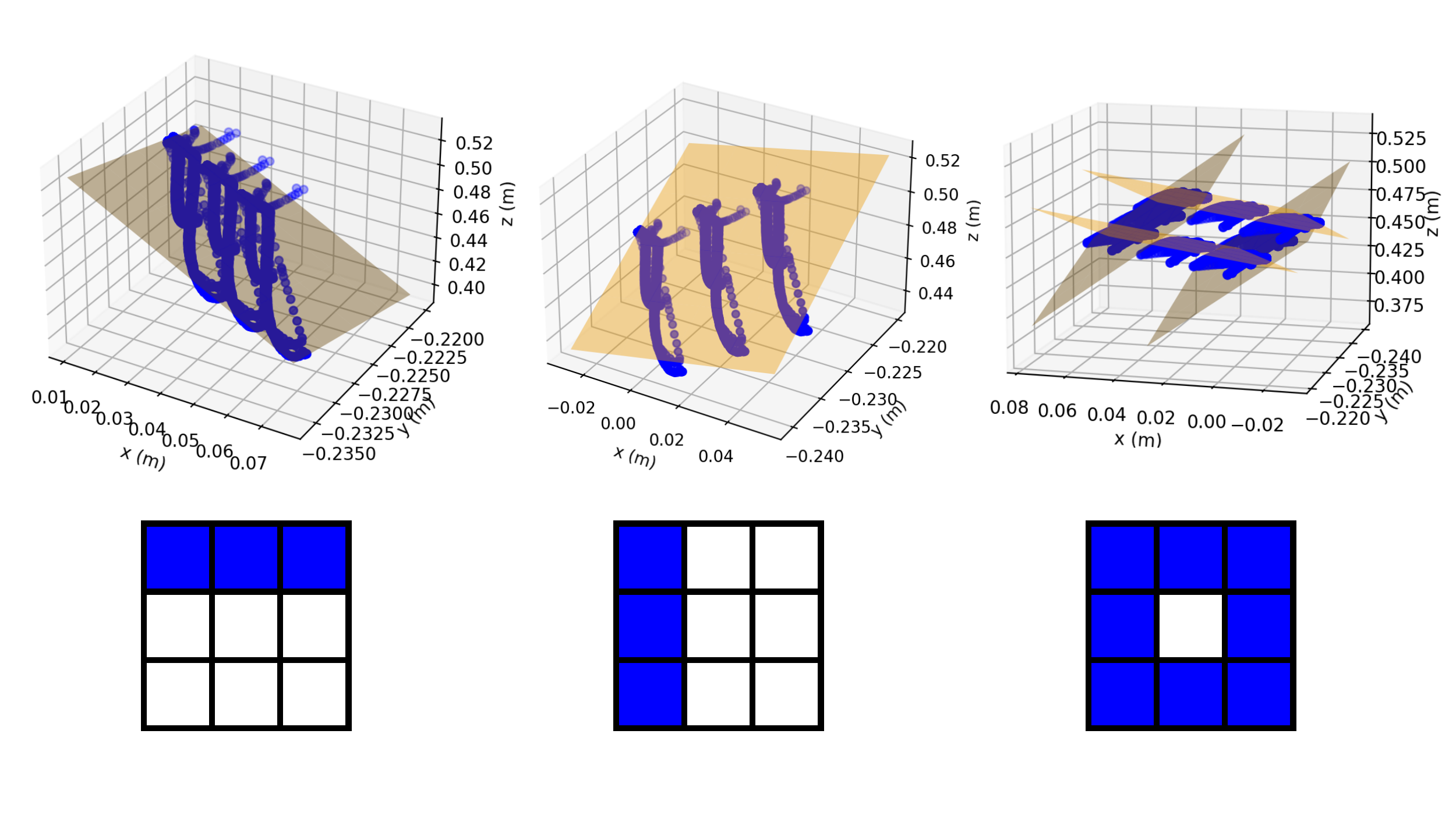}
   \vspace{-20pt}
  \caption{Point clouds created from transformations of the top (left), left (center), and all sides (right) of the $3 \times 3$ image. The planes are fitted with RANSAC on each side separately.}
  \label{vol_def_4_sides}
   \vspace{-20pt}
\end{figure}
To reduce the number of points in the point cloud, we treat each image as having dimensions of $3 \times 3$ pixels and scale them accordingly such that the pixels from the corners of the smaller image end up in the same coordinates as the corners of the original image. The point clouds for the 4 side faces of the volume are created by transforming 3 pixels from each image corresponding to the 4 sides of the image (see Figure \ref{vol_def_4_sides}). For the 2 base faces, the normal vector is first found by running RANSAC on the point cloud created by only performing the rotation transform on each image and omitting the translation. This plane represents the overall orientation of all the image frames, and is then translated to the coordinates of the start and end frames to close out the volume.

From these 6 planes, we calculate the relevant vertices by finding the intersections between the planes. The origin is chosen to be the intersection point closest to the top-left corner of the start frame. The vertex along the vector representing the new $z$ axis is the point closest to the top-left corner of the end frame, and the $x$ and $y$ vectors are chosen according to the right-hand rule.

\subsection{Volume Reconstruction}\label{sub:volumerecon}
Our reconstruction strategy uses the SDW algorithm from Huang et al.~\cite{HUANG2005153}, which assigns the intensity for each voxel in the volume by considering all the image pixels that fall into the spherical region within radius $r$ of the voxel. A voxel at coordinate $\vec{V}_C$ has intensity 
\begin{equation}\label{eq:voxel_intensity}
I(\vec{V}_C) = \frac{\sum_{k = 0}^n W_k I(\vec{V}_P^k)}{\sum_{k=0}^n W_k}
\end{equation}
where $I{(\vec{V}_P^k)}$ is the intensity of the $k$th pixel out of $n$ falling within the spherical region at coordinate $\vec{V}_P^k$.
$W_k$ is the weight of the $k$th pixel, defined by
\begin{equation}\label{eq:pixel_weight}
W_k = \frac{1}{(d_k + \alpha)^2}
\end{equation}
where $d_k$ is the distance between $\vec{V}_C$ and $\vec{V}_P^k$ and $\alpha$ is a positive parameter to adjust the reconstruction.

After the initial reconstruction using image pixel data, holes in the reconstructed volume are filled using the same algorithm, just with nearby voxels instead of image pixels. The hole filling involves multiple passes, with $r$ being increased after each pass up to a limit.

Implementation wise, we keep track of the numerators and denominators for each voxel separately and loop over each image frame, adding every pixel's contribution to nearby voxels. For each frame, the pixel vectors in the image are transformed to its coordinates in the volume by first applying the corresponding ROS transform to get coordinates in terms of the robot base, and then a second transform from the robot base coordinates to the volume coordinates as defined in the previous section. Specifically, for each 2D pixel vector $\vec{P}$, we have:
\begin{equation}\label{eq:pixel_transform}
\vec{V}_P = \begin{bmatrix} n_x & 0 & 0 \\ 0 & n_y & 0 \\ 0 & 0 & n_z \end{bmatrix}X^{-1}\begin{bmatrix} R & \vec{b} - \vec{u}\end{bmatrix}\begin{bmatrix} \vec{P} \\ 0 \\ 1 \end{bmatrix}
\end{equation}
where $R$ is the $3 \times 3$ rotation matrix and $\vec{b}$ is the $3 \times 1$ translation vector from the ultrasound image frame to the robot base frame, $\vec{u}$ is the new origin, $X$ is the $3 \times 3$ matrix with columns consisting of the new basis vectors, and $n_x,\,n_y,\,n_z$ are the number of voxels for each dimension of the volume. We augment the initial pixel vector to be able to perform matrix multiplication and vector addition with a single multiplication. Then, each pixel's distance and intensity to all voxels within a radius $r$ is calculated, and its contribution to each voxel's overall summation is added.

Once all the images have been processed, hole locations are determined by the voxels that have 0 for its denominator. These are temporarily set to 1, and each voxel's intensity is calculated by division. During the hole-filling stage, the algorithm is slightly modified to exclude contributions from voxels that are holes themselves.

All the raw ultrasound images were converted to gray scale for reconstruction, and each label class (e.g. vein, artery, and background) was reconstructed separately, with a pixel having a value of 1 if it was part of a label and 0 otherwise. The final label volume was created by assigning the label with the maximum value for each voxel.

\subsection{Volume Slicing}

From the reconstructed volumes, we now extract the augmented ultrasound images and labels.  To achieve this, the volumes are rotated around the $x$ and $y$ axes and fit within a new output volume using spline interpolation. The output volume is then expanded to include all the data from the original volume, with missing data filled with zeros.

Then, vertical slices are taken along the $z$ axis of the rotated output volume to generate the augmented images. These vertical slices of the rotated volume correspond to oblique slices of the original volume. Each augmented image slice is first checked to ensure that the percentage of pixels corresponding to the original volume (i.e., nonzero pixels) meets a certain threshold to ensure the augmented image contains enough useful information. If threshold is met, the maximum amount of zero pixels are cropped from each side of the image, and then the image is resized to $256 \times 256$ pixel grid. For our results, the volumes were rotated $\pm 30^{\circ}$ around both axes with $10^{\circ}$ steps. Vertical slices were taken once every 5 voxels, and our threshold was 40\% for the proportion of nonzero pixels needed in the augmented image.

\begin{figure}[tp]
  \centering
  \includegraphics[scale = 0.235]{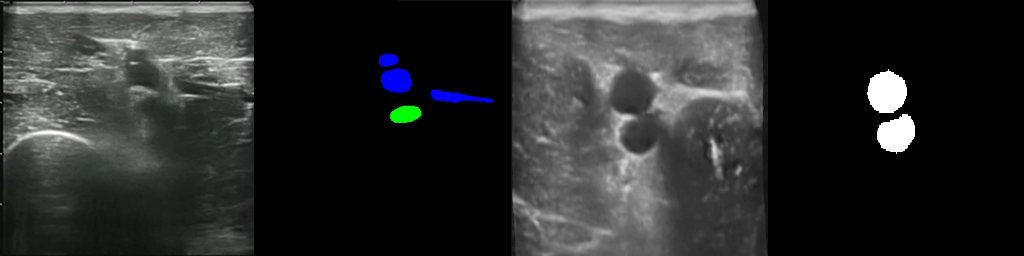}
  \vspace{-5pt}
  \caption{Raw ultrasound image and label (left) and sliced augmented image and label (right)}
  \label{vol_def_4_sides}
  \vspace{-20pt}
\end{figure}

\subsection{Training}
Our goal is to develop semantic segmentation of the femoral artery and vein in ultrasound images for the purpose of 3D reconstruction of the vessels for the robot to find an adequate needle insertion point. To that end, the deep learning network architecture that we use is a U-Net with an image classification model, ResNet34~\cite{7780459}, backbone in PyTorch~\cite{NEURIPS2019_9015} as the encoder using the Segmentation Models library \cite{Iakubovskii:2019}. ResNet34 was chosen due to its relatively high training speed and low memory demands~\cite{howard2018fastai}. ResNet34 consists of 34 layers with (3x3) convolutional filters using same padding, max-pooling layers and fully-connected layers, ending with a sigmoid activation function used to ascertain the output class. We train our network until the validation Dice loss converges with batch size 8 on lower resolution (256 × 256) images. We use the Adam optimizer\cite{DBLP:journals/corr/KingmaB14}
with a learning rate of 0.0001. The network was trained on NVIDIA RTX A6000 GPUs.

The model is evaluated using leave-one-subject-out crossvalidation protocol, in which each pig's data is used once as a test set while the remaining data forms the training set. Augmented images from RESUS are only used for training. To avoid over-fitting, images resliced from the test pig are removed from the training set. 

\subsection{Vessel Reconstruction}
Using the method described in \ref{sub:volumerecon}, we can input the predictions to reconstruct the vessels at the exact location they were scanned, thus creating an internal map of the vessels. To differentiate between different types of vessels (artery vs.\ vein), we use the knowledge of body's anatomy.

\section{EXPERIMENTS}
\label{sec:experiments}
We first test our method using the medical phantom, by creating a 3D volume and reslicing it. We consider two different models, one trained on unaugmented ultrasound images in transverse view, and the other trained on the unaugmented images along with the resliced images from the reconstructed volumes. We compare the models with two different test sets, the first consisting of the unaugmented images only, and the second consisting of both the unaugmented and the resliced images representing different variations in real world data, such as longitudinal vessels. We evaluate the results by reconstructing the vessels and comparing them qualitatively to the ground truth, as well as quantitatively with the IoU segmentation score. 

Then we transition to real world data, by training and testing on ultrasound data collected from live pigs. All the methods were tested using the ultrasound images gathered for each pig at the time of scanning. We compare our RESUS augmentation method with models trained on images collected from different subjects to assess the inter-subject generalization capabilities of the considered models. We also test models that were pretrained on ImageNet~\cite{imagenet}, and others that used common augmentation techniques for medical images such as horizontal and vertical flip, Gaussian noise addition, sharpening of the images, random brightness, and random contrast. Using the model predictions, we reconstruct the vessels and compare them visually with their ground truth. To corroborate our results, we use the IoU score metric to evaluate accuracy of segmentation. 

\section{RESULTS AND ANALYSIS}
\label{sec:results}
\begin{figure*}
  \centering
  \def\svgwidth{\textwidth}
  \includegraphics[width=0.99\textwidth]{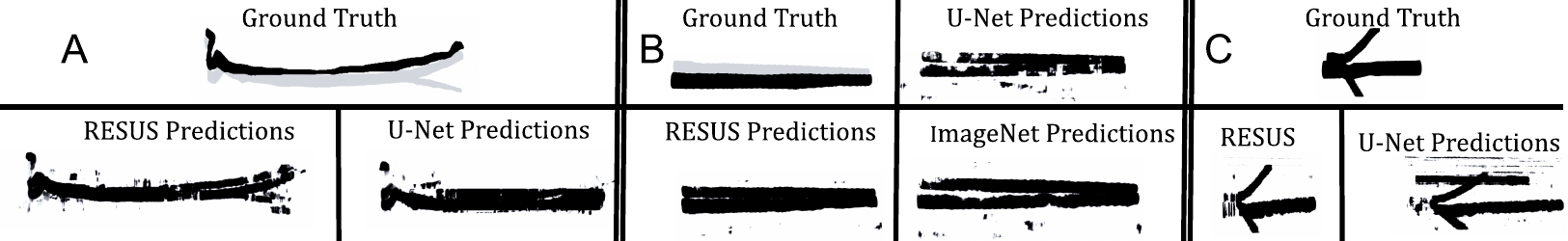}
  \vspace{-5pt}
  \caption{A) Pig 1 Vessel Reconstruction; B) Pig 4 Vessel Reconstruction;  C) Phantom Vessel Reconstruction}
  \label{reconstructed}
\end{figure*}

To evaluate the efficacy of our method we use qualitative and quantitative approaches. We observe that our method consistently surpasses the rest of the models at vessel reconstruction and removal of artifacts. As it can be observed in Figure \ref{reconstructed}C, RESUS reconstruction is visually closest to the ground truth. The U-Net prediction has artifacts that could be confused by a robot for a vessel, which could be life threatening in a real-life scenario. Quantitatively, the proposed method yields a 20\% improvement in IoU score for the phantom vessels. 

Evaluated on in-vivo animal data, the IoU score is on average 8\% higher than for the method using only unaugmented images and higher than attainable with other medical image segmentation techniques such as transfer learning with ImageNet (5\%) and random augmentations (23\%). Paired $t$-test performed to assess significance of the observed improvements, yields p-values of $0.034$, $0.049$ and $0.00046$, respectively. 
Thus, we can confidently conclude that the proposed approach prevalently improves accuracy of image segmentation.  Table \ref{pigresults} summarizes quantitative results.  

While the IoU scores might seem low, one of the limitations of this work includes many flaws in the labels which lead to lower than attainable success rates of the models. Another limitation was that for several pigs, the ultrasound was stationary, while in others it was zigzagging making it difficult to create a volume, thus for the eight pigs which we used to train on, we only had five volumes of small sections of the data. Lastly, the experiments schedule was sparse to allow for technical developments of the prime purpose of the animal study, some of which mildly affected the collected ultrasound data. Even with the aforementioned obstacles, by comparing all qualitative results as shown in Figure \ref{reconstructed}(A\&B) with the different methods, we can observe that the RESUS segmentation is again the closest to the ground truth. By including data obtained from our method to augment the training set, our vessel reconstruction results qualitatively better represent the shape and bifurcations of the vessels, and the resulting segmentations are more accurate than attainable with alternative approaches under consideration. We observe that these improvements can increase the likelihood of successful needle insertions for various kinds of treatments in trauma care in the field.




\setlength{\arrayrulewidth}{0.25mm}
\begin{table}[bp]
\caption{IoU score of Segmentations tested on the Phantom}
\vspace{-10pt}
\label{phantomresults}
\begin{center}
\begin{tabular}{c|cc}
\toprule[1pt]
 & \textbf{Train US} &  \textbf{Train US and RESUS} \\ \midrule
\textbf{Test US} & 0.536 & \textbf{0.731}\\
\textbf{Test US and RESUS} & 0.186 & \textbf{0.779} \\ \bottomrule[1pt]
\hline
\end{tabular}
\end{center}
\end{table}

\setlength{\arrayrulewidth}{0.25mm}
\begin{table}
\caption{IoU Scores of Segmentations Tested on the pigs}
\vspace{-10pt}
\label{pigresults}
\begin{center}
\begin{tabular}{l|ccccc}
\toprule[1pt]
 & \textbf{U-Net} & \textbf{ImageNet} & \textbf{Aug} & \textbf{RESUS}\\
\midrule
\textbf{Pig 1 ($n=850$)} & 0.3685 & 0.591 & 0.383 & \textbf{0.610}\\
\textbf{Pig 2 ($n=951$)} & 0.587 & \textbf{0.607} & 0.537 & 0.603\\
\textbf{Pig 3 ($n=1307$)} & 0.631 & 0.691 & 0.324 & \textbf{0.693}\\
\textbf{Pig 4 ($n=900$)} &0.513 & 0.590 & 0.350 & \textbf{0.665}\\
\textbf{Pig 5 ($n=900$)} & \textbf{0.665} & 0.408 & 0.319 & 0.610\\
\textbf{Pig 6 ($n=500$)} & 0.7483 & \textbf{0.785} & 0.536 & 0.7529 \\
\textbf{Pig 7 ($n=2070$)} & 0.422 & 0.509 & 0.262 & \textbf{0.571}\\
\textbf{Pig 8 ($n=1200$)} & 0.574 & 0.531 & 0.555 & \textbf{0.601}\\
\midrule
\textbf{AVERAGE} & 0.557 & 0.592 & 0.408 & \textbf{0.638}\\ \bottomrule[1pt]
\end{tabular}
\end{center}
\vspace{-15pt}
\end{table}

\section{CONCLUSION}
\label{sec:conclusion}
This paper introduces RESUS, an augmentation method for image segmentation, specific to imaging modalities consisting of multiple 2D slices of a 3D volume. Examples of such modalities include ultrasound data and CT scans. We have demonstrated RESUS on in-vivo animal data, showing performance improvements over models trained on unaugmented data alone, models trained on ImageNet, and models trained using commonly used medical imaging augmentation techniques. 

The goal of RESUS is to support using ultrasound imaging as a sensor for robotic control. Currently, intelligent use of ultrasound imaging is limited by the required large volumes of training data and the cost of harvesting labels from domain experts. RESUS directly addresses these challenges by providing affordable image augmentations, specific to ultrasound images that are slices of 3D volumes. It yields high quality ultrasound image segmentation, which when used for robotic control, can automate important tasks in critical and trauma care. 

Currently, we use the anatomical knowledge of the body \cite{Chun2018-eq} to differentiate between the femoral artery and vein. 
Next, we will leverage models of deformation to distinguish between veins and arteries automatically, and integrate our tool with the needle insertion controller.
We also observe that even if the IoU score of our model was not always the highest, our method showed the qualitatively best vessel reconstruction consistently. Thus, the IoU score may not be the best evaluation metric for the segmentation of our images, and we will look for a better one. 
Finally, the current robotic setup is not yet suitable for the application in field emergency care: form factor, weight, power needs, sterility, safety and FDA certification, all need to be addressed to facilitate real-world deployment. 

\addtolength{\textheight}{-10cm}   

\section*{ACKNOWLEDGMENTS}
We would like to thank Nico Zevallos, Dr.\ Michael R.\ Pinsky, and Dr.\ Hernando Gomez for gathering the data for our experiments, and Mononito Goswami for thoughtful suggestions on this manuscript. 


\bibliographystyle{IEEEtran}
\bibliography{tracir}

\end{document}